%% file: article_arXiv.tex
\documentclass[a4paper]{spie}
\usepackage[utf8]{inputenc}
\usepackage[T1]{fontenc}
\usepackage[american,english]{babel}
\usepackage{lmodern}
\usepackage{cmap}
\usepackage{graphicx}
\usepackage{xcolor}
\usepackage{siunitx}
\usepackage{amsmath}
\usepackage{amssymb}
\usepackage{mathtools}
\usepackage{float}
\graphicspath{{img/},{../img/}}
\usepackage{subfigure}

\usepackage{transparent}
\usepackage{pgf}
\usepackage{tikz}

\title{A novel high-contrast imaging technique based on optical tunneling to search for faint companions around bright stars at the limit of diffraction} 

\author{Dominik Derigs\supit{a}, Lucas Labadie\supit{a}, Dhriti Sundar Ghosh\supit{b} and Laëtitia Abel-Tibérini\supit{c}, 
\skiplinehalf
\supit{a}I. Physikalisches Institut, Universität zu Köln, Zülpicher Straße 77, 50937 Köln, Germany; \\
\supit{b}ICFO - Institut de Ciències Fotòniques, Barcelona, Spain; \\
\supit{c}Institut Fresnel, Marseille Cedex, France
}

\authorinfo{Responsible author: D.Derigs\\\indent E-mail: derigs@ph1.uni-koeln.de}

\let\oldthepage\thepage
\renewcommand*\thepage{\oldthepage}
\begin{document} 
\maketitle 

\begin{abstract}
We present a novel application of optical tunneling in the context of high-angular resolution, high-contrast techniques with the aim of improving direct imaging capabilities of faint companions in the vicinity of bright stars. In contrast to existing techniques like coronagraphy, we apply well-established techniques from integrated optics to exclusively extinct a very narrow angular direction coming from the sky. This extinction is achieved in the pupil plane and does not suffer from diffraction pattern residuals. We give a comprehensive presentation of the underlying theory as well as first laboratory results.
\end{abstract}

\keywords{optical tunneling, integrated optics, high-angular resolution, high-contrast, direct imaging, exoplanet detection}

\section{INTRODUCTION}
\label{sec:intro}
Opening up the area of high-angular resolution is scientifically appealing as the search and direct imaging of faint companions in the vicinity of bright stars -- ranging from cool brown dwarfs down to extrasolar planets -- still remains very challenging. It has long been a major area of research in astrophysics.

Direct imaging of faint companions is a high-angular resolution and meanwhile high-contrast problem, classically solved with coronagraphic systems, which are based on the usage of an opaque or semi-opaque mask placed in the image plane at the position of the primary star. The corollary to direct imaging is direct spectroscopy, which permits the determination of physical properties like the atmospheric compositions, temperatures, and surface gravity of the companions. 
However, imaging extra-solar planets by means of spectro-photometry of their atmospheres is very challenging, and the practical hurdles, that have been described by Mawet D.\ et al.\cite{mawet2012review}, are numerous. Most important:
\begin{description}
	\item[The angular separation between planets and stars is generally very small.]\hfill\\
	The small angular separation ($< 0.1^{\prime\prime}$ for a $\SI{1}{AU}$ type planet at a distance of $\SI{10}{pc}$) requires diffraction limited capabilities on ground-based telescopes in the visible.
	\item[The contrast might be very low.]\hfill\\
	The contrast between a planet and its host star ranges from $10^{-3}$ (giant hot planets, IR) to $10^{-10}$ (Earth-like planets, visible).
\end{description}
Research fields like extra-solar biology rely on emission spectra measurements from chemical markers like water, CO${}_2$ or ozone that can only be observed by means of direct spectroscopy. Therefore, high-contrast direct spectroscopy of extra-solar planetary systems might get the most valuable toolkit for the characterization of planets and their host systems in the future. This implies that the glare of the parent star has to be separated from the light that comes from the companion itself. In order to achieve such a separation, one has to unite high-angular resolution and high-contrast techniques.

In this paper we present, for the first time to our knowledge, a novel application of optical tunneling.
The extinction of the central star is achieved directly in the pupil plane, rather than in the image plane. We use the well-known prism coupler technique\cite{Tien:69,Tien:71,Labadie:M-Lines,FTIR_paper,Saleh_Teich:Fundamentals,Ulrich:73,Ulrich:71} to couple a specific angular direction of the incident light into a juxtaposed single-mode planar waveguide.
The incident light, coming from the telescope, is totally internally reflected inside the prism. Due to the consistency condition of single-mode waveguides, one specific angular direction of the incident beam (in our case that of the host star) gets coupled exclusively into the waveguide and is therefore highly attenuated at the detector site. Meanwhile, all other angular directions of the light (including the light from an off-axis companion) are not altered and can be detected at the output facet of the prism.

We present here the underlying theory of this concept – including its wavelength dependence – as well as the first results of a lab-experiment showing that 95\% extinction can be achieved routinely.

\section{THEORETICAL EVALUATION} 
The amplitude of an electromagnetic wave that is totally internally reflected at an interface is nonzero behind it, but falls off exponentially, allowing the extraction of light from waveguides where it is confined by total internal reflection at both interfaces. This can be compared to the well-known tunneling effect in quantum-mechanics where a particle could be present in a classically forbidden region, where the probability of presence is similarly decaying exponentially behind the interface. Due to this analogy, the phenomenon we are going to discuss is also called \emph{optical tunneling}. It cannot be explained in terms of geometrical optics so that one has to examine the electromagnetic fundamentals of total internal reflection itself.

\subsection{Continuity Conditions at an Interface}\label{scn:Theo:Reflection_and_Refraction}
When the wave approaches an interface where the refractive index changes, one has to consider the continuity conditions resulting from Maxwell's equations.
They require that the components of the electric field have to be continuous at the interface and cannot suddenly fall off to zero.
The tangential component has to be continuous at the interface. The normal component has a jump but is not of further interest for our discussion.
\begin{figure}[b]
	\begin{center}
		\includegraphics[height=7cm]{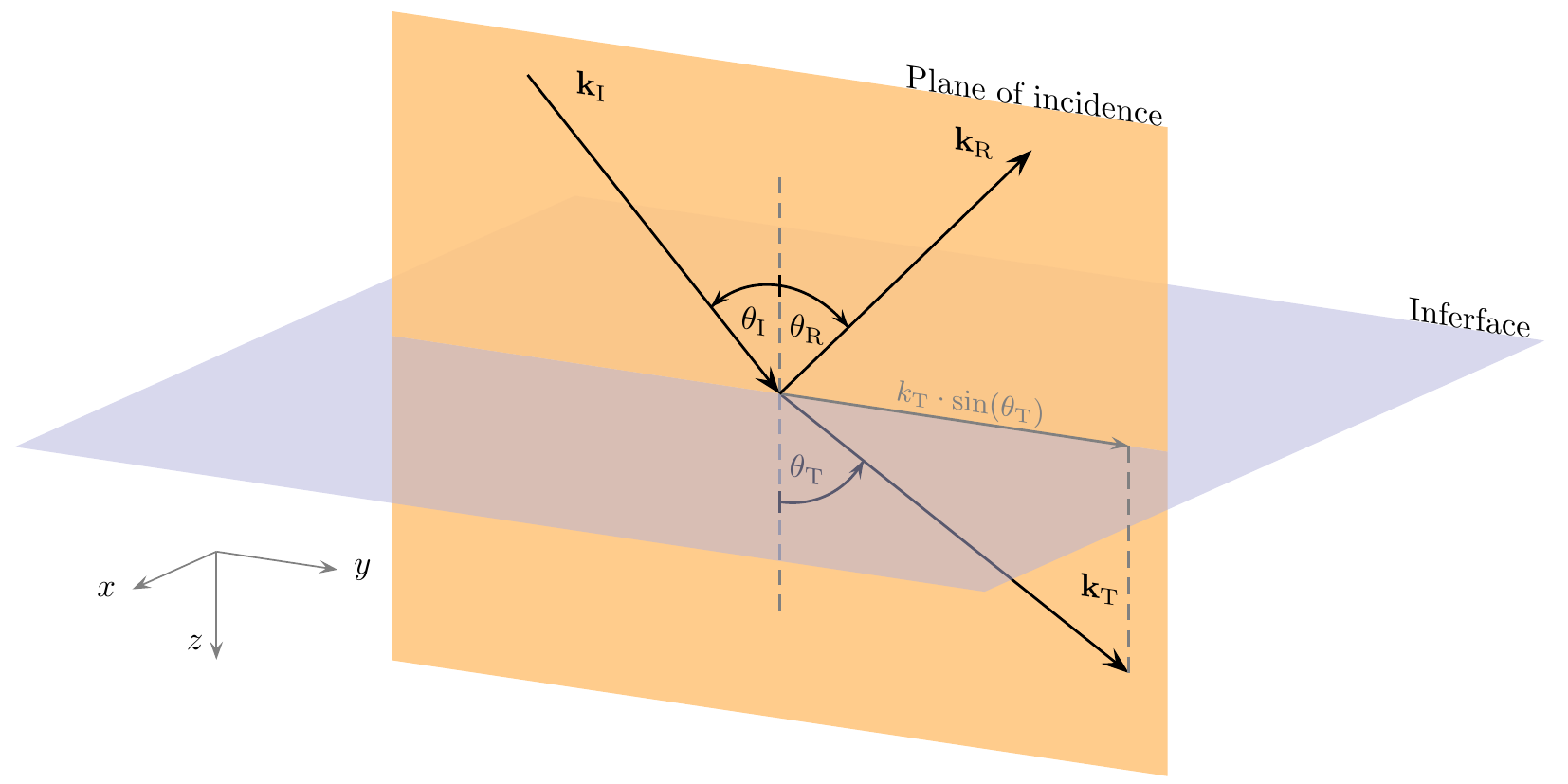}
	\end{center}
	\caption[example]
	{ \label{fig:Theo:Reflection-and-Refraction-3D} 
The incident beam hits the interface on the intersection line of the plane of incidence and the interface. $\mathbf k_\mathrm{I}$ marks the incident beam (input), $\mathbf k_\mathrm{R}$ marks the reflected beam (output), and $\mathbf k_\mathrm{T}$ marks the beam which is transmitted in the other medium.}
\end{figure}
\begin{subequations}\begin{align}\label{eq:Theo:continuity_condition}
	\left(\mathbf k_\mathrm{I} - \mathbf k_\mathrm{T} \right) \cdot \mathbf{r} &= 0 \ ,
\shortintertext{and}
	\left(\mathbf k_\mathrm{I} - \mathbf k_\mathrm{R} \right) \cdot \mathbf{r} &= 0 \ ,
\end{align}\end{subequations}
where $\mathbf r$ is the unit vector normal to the interface-plane. $\mathbf k_i$ are the wavevectors whose subscripts $\mathrm{I}$, $\mathrm{T}$, and $\mathrm{R}$ represent the incident, the transmitted, and the reflected wave, respectively. In order to simplify the following analysis, 
we will assume that the interface lies in the $x$-$y$-plane (cf.~Fig.~\ref{fig:Theo:Reflection-and-Refraction-3D}). Thus, all three wavevectors are in the same plane, the plane of incidence, which is the $y$-$z$-plane in our case. Note that this is no limitation, but only a simplification of notation. 

By applying the inner product in equations (\ref{eq:Theo:continuity_condition}/b) one finds that the vector components of the wavevectors parallel to the interface have to be equal:
\begin{subequations}\begin{align}\label{eq:Theo:y-components_of_wavevector}
	k_{\mathrm{I},y} - k_{\mathrm{T},y} = 0 \quad \Rightarrow \quad k_{\mathrm{I},y} &= k_{\mathrm{T},y} \\
	k_{\mathrm{I},y} - k_{\mathrm{R},y} = 0 \quad \Rightarrow \quad k_{\mathrm{I},y} &= k_{\mathrm{R},y} \ ,
\end{align}\end{subequations}
showing that the continuity conditions require the wavevectors and therefore the electromagnetic fields to be present on either side of the interface.

By combining the norm of the two-dimensional wavevector, $k_\mathrm{T}^2 = k_\mathrm{T,z}^2 + k_\mathrm{T,y}^2$, with the law of reflection, $n_1 \cdot \sin(\theta_\mathrm{I}) = n_2 \cdot \sin(\theta_\mathrm{T})$, and the linear dispersion relation of planar waves, $k_\mathrm{T} = n_\mathrm{T} \cdot (\omega/c_0)$, we get to the expression
\begin{equation}
	k_{\mathrm{T},z}^2 = k_\mathrm{T}^2 - k_\mathrm{T,y}^2 =  k_\mathrm{T}^2 \cdot \left(1 - \sin^2(\theta_\mathrm{T})\right) = \left(n_\mathrm{T} \cdot \frac{\omega}{c_0}\right)^2 \cdot \left(1-\frac{n_\mathrm{I}^2}{n_\mathrm{T}^2} \cdot \sin^2 \left(\theta_\mathrm{I}\right)\right) \ ,
\end{equation}
where $n$, $\omega$, and $c_0$ denote the refractive index of the medium the wave is propagating in, the angular frequency of the wave, and the speed of light in vacuum, respectively.

\begin{figure}[b]
	\begin{center}
		\begin{tabular}{c}
			\includegraphics[height=7cm]{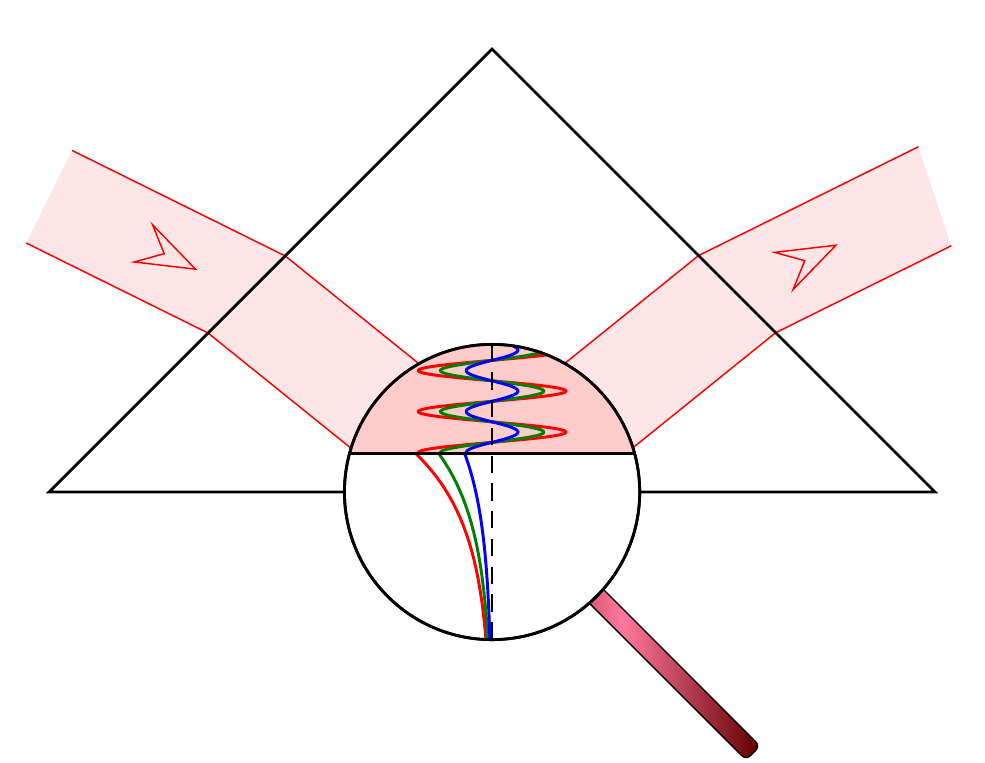}
		\end{tabular}
	\end{center}
	\caption[example]
	{ \label{fig:Theo:Standing_Wave} 
Snapshot of the standing wave pattern at three different times. Although the amplitude of the field changes in time, the positions of the wave nodes remain unchanged.}
\end{figure}

For $(n_\mathrm{I} / n_\mathrm{T}) \cdot \sin \left(\theta_\mathrm{I}\right) > 1$ the wave cannot propagate freely into the material as $k_{\mathrm{T},z}$ becomes purely imaginary. We can rewrite the wavevector as
\begin{equation}\label{eq:beta}
	k_{\mathrm{T},z} = \pm i \beta \cdot k_\mathrm{T} \ ,
\end{equation}
with $\beta^2 = (n_\mathrm{I}^2 / n_\mathrm{T}^2) \cdot \sin^2 \left(\theta_\mathrm{I}\right) - 1$ we introduce the \emph{propagation constant}. The critical angle for total internal reflection at the interface is given by $\theta_\mathrm{crit} = \arcsin\left( n_\mathrm{T} / n_\mathrm{I} \right)$.
The exponentially decaying field penetrating the second medium (refer to Fig.~\ref{fig:Theo:Standing_Wave}) is called the \emph{evanescent field}. It has first been discovered by Sir Isaac Newton over three hundred years ago~(1704).

The wavefunction of this transmitted electric field is
\begin{equation}\label{eq:Theo:Evanescent_Electric_Field}
	\mathbf E_\mathrm{T}(\mathbf{r}) = \mathbf E_{\mathrm{T},0} \cdot e^{i \left(\mathbf k_{\mathrm{T}} \cdot \mathbf r - \omega t \right)} \stackrel{\eqref{eq:beta}}{=} \mathbf E_{\mathrm{T},0} \cdot e^{ - \beta k_{\mathrm{T}} \cdot z} \cdot e^{i \left( k_{\mathrm{T},y} \cdot y - \omega t \right)} .
\end{equation}
At first glance, this wavefunction seems to be exactly the same as for a normal wave propagating in $z$ and $y$ direction and oscillating in time. But since $k_{\mathrm{T},z}$ is purely imaginary, this equation is no longer describing a propagation in $z$-direction. Instead, it describes a wave whose amplitude drops off exponentially as it penetrates into the material, being highest at the interface and becoming negligible close behind.

The point at which the amplitude of the evanescent field decays to its $1/e$ value is called the \emph{penetration depth} of the evanescent field. It can be derived from \eqref{eq:Theo:Evanescent_Electric_Field}, using $k = 2 \pi / \lambda$,
\begin{equation}\label{eq:penetration_depth}
	z_{1/e} = \frac{1}{k \beta} = \lambda \cdot \left( 2 \pi \cdot \sqrt{\frac{n_\mathrm{I}^2}{n_\mathrm{T}^2} \cdot \sin^2(\theta_\mathrm{I}) - 1} \right)^{-1} .
\end{equation}
The penetration depth is a function of the ratio between both refractive indices $n_\mathrm{I} / n_\mathrm{T}$ and the angle of incidence $\theta_\mathrm{I}$ and is defined only above the critical angle for total internal reflection (it is purely imaginary below). It is interesting to note that the penetration depth decreases from infinite value (at the critical angle) down to its minimum value at grazing incidence ($\theta_I = \SI{90}{\degree}$). The evanescent waves that are created by beams that are reflected sufficiently close to the critical angle have penetration depths that are surpassing any upper boundary, thus smoothly transiting to the non-evanescent regime where the wave can travel inside the other medium and therefore can reach infinite distances trivially. Apart the critical angle the penetration depth decreases quickly. The penetration depth is an important measure needed when setups that rely on evanescent fields are to be designed. Its value is typically a fraction of a wavelength.

When a medium with high refractive index is placed in very close proximity to another medium with comparable high refractive index and when they are separated by a medium with low refractive index, e.g.\ air, parts of the evanescent wave can be coupled between both high refractive index media.
A wave is excited in the second medium. The reflected wave in the initial medium is attenuated thereupon and the energy is transferred into the new medium. The \emph{coupling efficiency}, which is the ratio of coupled to total intensity, is very sensitive to the gap thickness which should be of the same order as the penetration depth given by \eqref{eq:penetration_depth}.


\subsection{Planar Slab Waveguides} 
\begin{figure}[h]
	\begin{center}
		\begin{tabular}{c}
			\includegraphics[height=5cm]{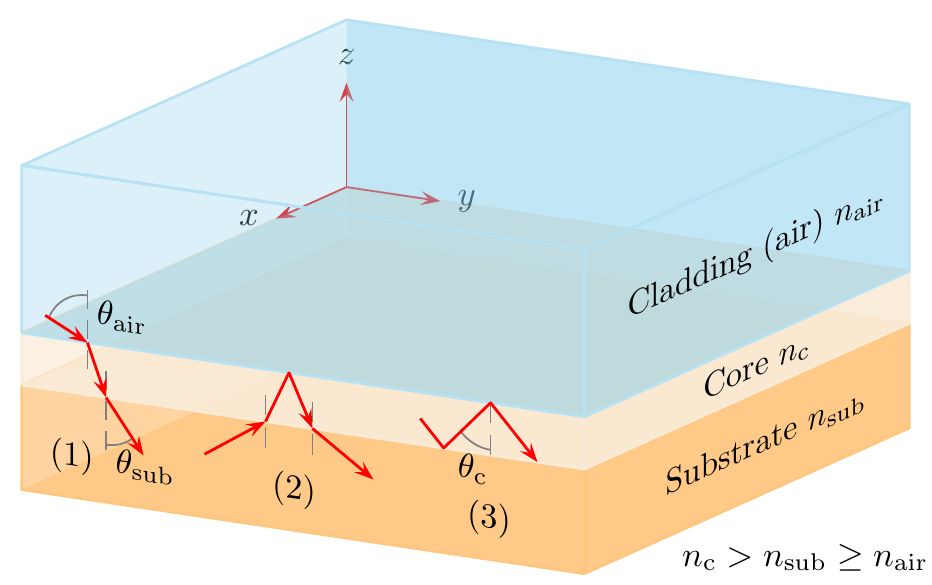}
		\end{tabular}
	\end{center}
	\caption[example] 
	{ \label{fig:Planar_slab_waveguide} 
A planar slab waveguide as described in the text.}
\end{figure}

The planar slab waveguide shown in Fig.~\ref{fig:Planar_slab_waveguide} is the simplest dielectric waveguide. A planar film (called \emph{core}, refractive index $n_\mathrm{c}$) is deposited on a substrate with lower refractive index ($n_\mathrm{sub}$) and covered by a cladding layer having an even lower refractive index ($n_\mathrm{cladd}$), in general. In our case, it will be assumed that the cladding material is, for simplicity, the surrounding air itself, in which case $n_\mathrm{cladd} = n_\mathrm{air} \approx 1$. The light is confined by total internal reflection at the core-substrate and core-cladding interfaces. The substrate takes a twin role: on one hand, it provides an interface for the total internal reflection between itself and the core layer, on the other hand, it is necessary in terms of mechanical stability, since the core layer is often only a few microns (or even less) in thickness.

The picture of light guidance in the slab waveguide is one of light rays tracing a zig-zag path in the thin core layer, traveling up and down, since total internal reflection will happen at the core boundaries. The light propagates in $y$-direction, while the confinement occurs transversely in $z$-direction. This translates into a wave propagating parallel to the film with constant speed. Therefore, it seems natural to introduce a propagation constant
\begin{equation}
	\tilde \beta := \frac{\omega}{v_\mathrm{ph}} = k n_\mathrm{c} \sin(\theta_\mathrm{c}) \ .
\end{equation}

Due to the critical angles of total internal reflection at the core interfaces, the boundaries of the propagation constant are given by $k \cdot n_\mathrm{sub} < \tilde \beta < k \cdot n_\mathrm{c}$. One often uses the so-called \emph{effective refractive index} having an analogous meaning like the ordinary refractive index increasing the wavenumber (phase change per unit length) in homogeneous transparent media. It is defined by
\begin{equation} \label{eq:Theo:eff_ref_index}
	n_\mathrm{eff} = \tilde \beta / k = n_\mathrm{c} \sin(\theta_\mathrm{c}) ,
\end{equation}
and is bound similar to the propagation constant by $n_\mathrm{sub} < n_\mathrm{eff} < n_\mathrm{c}$.

One can find three possible types of propagation (see Fig.~\ref{fig:Planar_slab_waveguide}): Case (1) \emph{Air-Substrate modes} where small core-interface angles $\theta_\mathrm{c}$ prevent internal reflections at any surface, Case (2) \emph{Substrate modes} where larger angles allow total internal reflections at the air-core interface only, and Case (3) where even larger angles of incidence, i.e.\ $\theta_\mathrm{c} > \arcsin(n_\mathrm{sub} / n_\mathrm{c})$ lead to a situation where the light wave cannot escape the core layer anymore and will be totally internally reflected at both the upper and the lower interfaces of the core layer. The light is trapped resp.~\emph{guided} inside the core.
\subsection{The Self-Consistency Condition} 
Due to the requirement that the wave has to interfere constructively with itself, it must obey a phase shift of $2\pi \cdot m$, where $m$ is an integer ($m = 0,1,2,\dots$) identifying the mode number. Otherwise -- even if there would be only a very tiny difference in the phase shift -- this difference would sum up and eventually the wave would extinguish itself due to destructive interference.

\noindent There are four phase shifts in the zig-zag path, that need to be taken into account:
\begin{enumerate}
	\item Starting from the core-substrate interface, the beam has to travel a distance $d \cdot \cos(\theta_\mathrm{c})$ to reach the core-air interface (core thickness $d$).
	\item When totally internally reflected at the dielectric boundary, a phase shift of $- \Phi_\mathrm{core,cladd}$ is introduced.
	\item Going down again to the core-substrate interface, the beam has to travel the same direction $d \cdot \cos(\theta_\mathrm{c})$.
	\item It will encounter a phase shift due to the second total internal reflection, $- \Phi_\mathrm{core,sub}$.
\end{enumerate}

\noindent These parts form the so-called \emph{self-consistency condition}
\begin{equation}\label{eq:Theo:self_consistency_condition}
	2 \, k \, n_\mathrm{c} \, d \cdot \cos(\theta_\mathrm{c}) - \Phi_\mathrm{core,cladd} - \Phi_\mathrm{core,sub} = 2 \pi \cdot m \ ,
\end{equation}
which is essentially the dispersion equation of the guided modes. The mentioned phase shifts are depending on the angle of incidence $\theta_\mathrm{c}$ on the interface and can be extracted from the Fresnel relations\cite{Saleh_Teich:Fundamentals}:
\begin{subequations}
	\begin{align} \label{eq:Theo:Phi_TE}
		\tan (\Phi_\mathrm{TE}/2) &= \frac{\sqrt{\sin^2 \theta_\mathrm{c} - n_B^2 / n_A^2}}{\cos \theta_\mathrm{c}} \, \\
		\tan (\Phi_\mathrm{TM}/2) &= \frac{n_A^2}{n_B^2} \cdot \frac{\sqrt{\sin^2 \theta_\mathrm{c} - n_B^2 / n_A^2}}{\cos \theta_\mathrm{c}} \ . \label{eq:Theo:Phi_TM}
	\end{align}
\end{subequations}
Here, the phase shifts represent, in fact, the Goos-Hänchen shifts\cite{GoosHaenchen} appearing at total internal reflection on dielectric interfaces. Of course, the appropriate substitutions for the refractive indices $n_\mathrm{A}$ and $n_\mathrm{B}$ must be applied ($n_\mathrm{A} > n_\mathrm{B}$) prior to the usage of these formulas. This polarization dependence is the reason why TE and TM modes have different dispersion relations.

\subsection{The Mode Dispersion Equation}
\begin{figure}[b]
	\centering %
	\def\svgwidth{0.8\textwidth} %
	\def\wavelengthA{$d = \SI{250}{nm}$}
	\def\wavelengthB{$d = \SI{25}{nm}$}
	\def\wavelengthC{$d = \SI{500}{nm}$}
	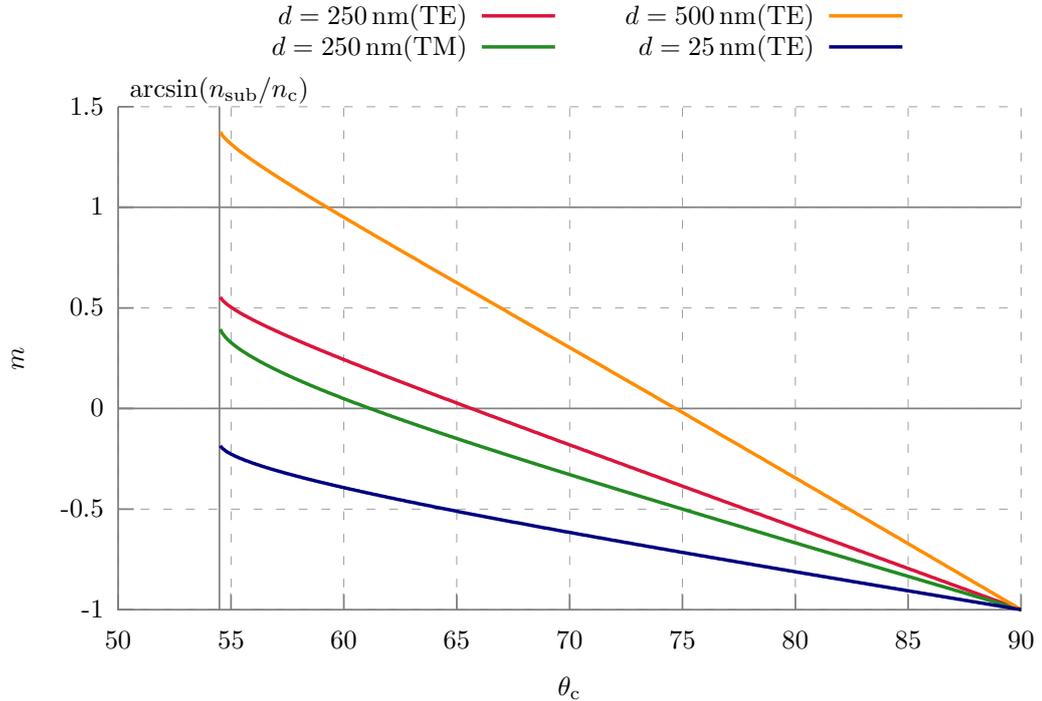
	\caption{ \label{fig:modedispeq}
	Plot of \eqref{eq:Theo:mode_disp_eq} for different film layer thicknesses $d$. $\lambda = \SI{632.8}{nm}$, $n_\mathrm{c} = 1.79$, $n_\mathrm{sub} = 1.457$}
\end{figure}

By combining \eqref{eq:Theo:self_consistency_condition} and (\ref{eq:Theo:Phi_TE}/b) one gets the important \emph{mode dispersion equation} which describes the behavior of the modes in the waveguide
\begin{align}
	m(\theta_c) = \left[\frac{4 \pi}{\lambda} \, n_\mathrm{c} \, d \cdot \cos(\theta_\mathrm{c}) - 2 \arctan\left( \alpha_1 \frac{\sqrt{\sin^2 \theta_\mathrm{c} - n_\mathrm{cladd}^2 / n_\mathrm{c}^2}}{\cos \theta_\mathrm{c}}\right) - 2 \arctan\left( \alpha_2 \frac{\sqrt{\sin^2 \theta_\mathrm{c} - n_\mathrm{sub}^2 / n_\mathrm{c}^2}}{\cos \theta_\mathrm{c}}\right)\right]/(2 \pi) \label{eq:Theo:mode_disp_eq}
\end{align}
with the polarization dependent constants $\alpha_1$ and $\alpha_2$ which are defined by
\begin{subequations}
	\begin{align}
		\alpha_1 &:= \begin{cases} \label{eq:Theo:alpha1}
			1 \qquad\qquad\qquad	&	\mbox{TE polarization} \\
			n_\mathrm{c}^2 / n_\mathrm{cladd}^2 &	\mbox{TM polarization} \\
	\end{cases} \\
		\alpha_2 &:= \begin{cases} \label{eq:Theo:alpha2}
			1 \qquad\qquad\qquad	&	\mbox{TE polarization} \\
			n_\mathrm{c}^2 / n_\mathrm{sub}^2 &	\mbox{TM polarization} \ . \\
	\end{cases}
	\end{align}
\end{subequations}
The mode number $m$ is an integer, starting with zero identifying the \emph{fundamental mode}. To get the allowed bounce angles, one has to solve \eqref{eq:Theo:mode_disp_eq} for positive integer solutions $m(\theta_\mathrm{c}) = 0,1,2,\dots$ which can only be done numerically (see also Fig.~\ref{fig:modedispeq}). As $d/\lambda$ gets smaller, the zig-zag path gets steeper ($\theta_\mathrm{c}$ gets smaller). If there is no solution for a given mode $m$, this mode is not supported. Therefore, not all angles are allowed but only a characteristic, discrete set of modes is possible.
The absolute number of modes can be concluded from \eqref{eq:Theo:mode_disp_eq} by setting $\theta_\mathrm{c} = \theta_\mathrm{crit}$:
\begin{equation}
	M = 1 + \lfloor m(\theta_\mathrm{crit}) \rfloor ,
\end{equation}
where $\lfloor x \rfloor$ refers to the largest integer not greater than $x$ (e.g.~$\lfloor 0.8 \rfloor = 0$ and $\lfloor -0.2 \rfloor = -1$).

\section{The prism coupler}
One of the basic components of integrated optics are beam couplers that can couple the energy of a free-space or a guided light beam into one excitation mode of a waveguide.
The first laboratory experiments that proved highly efficient coupling to a thin film using a coupling prism have been described in the pioneering work of Tien et al.\cite{Tien:69} in 1969. They reported that laser beams can be coupled efficiently in (and out of) waveguides through excitation of the guiding modes using a prism coupler.
In this investigation we have used the prism coupler, as high efficiently coupling has already been reported in previous works. Nevertheless, there are other waveguide couplers like the grating coupler introduced by Dakss et al.\cite{grating}. They achieved a coupling efficiency of $\approx 40\%$.
 
 \begin{figure}[b]
 	\begin{center}
 		\begin{tabular}{c}
 			\includegraphics[scale=1]{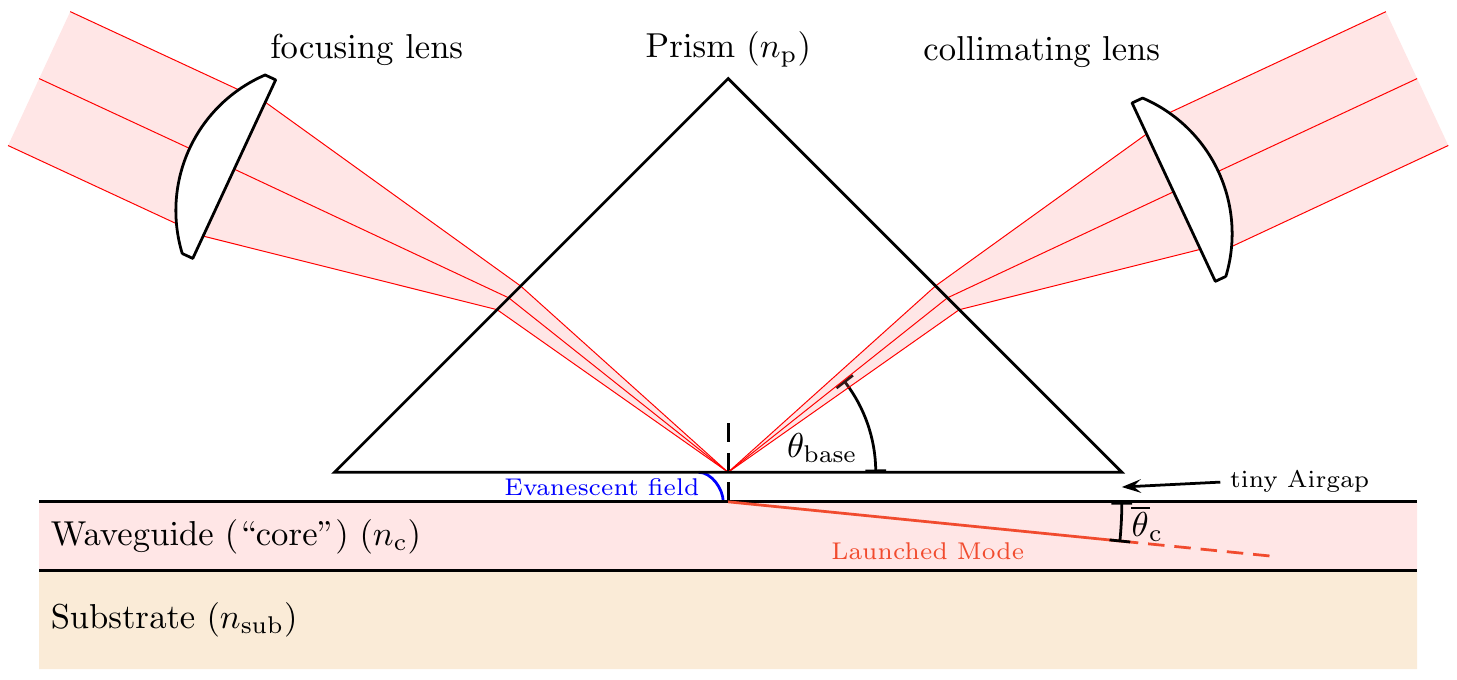}
 		\end{tabular}
 	\end{center}
 	\caption[example]
 	{ \label{fig:Displacement} 
 	The prism coupler can be used to exclusively excite one specific mode of the waveguide. $\theta_\mathrm{base}$ is the angle of incidence respectively the angle of reflection at base of the prism.}
 \end{figure}
 
We have placed a waveguide parallel to the base of the prism (refer to Fig.~\ref{fig:Displacement}). To achieve an adjustable and continuous spacing between prism and waveguide, they are pushed towards each other using a micrometer drive.
As has been shown in section \ref{scn:Theo:Reflection_and_Refraction}, the fields above and below the interface on the prism base must have the same horizontal wave motion.
The forming evanescent fields penetrate each other and constitute the coupling. The parts of the fields that overlap are common to the prism and the waveguide and couple power exclusively into one single mode of the waveguide.
We can rewrite \eqref{eq:Theo:y-components_of_wavevector} to the \emph{phase-matching condition}:
\begin{equation} \label{eq:Theo:phase-matching_condition}
	n_\mathrm{p} \cdot \cos(\theta_{\mathrm{base},m}) = n_\mathrm{c} \cdot \cos(\overline \theta_{\mathrm{c},m}) .
\end{equation}

In order to be able to excite all possible waveguide modes, the refractive index of the prism, $n_\mathrm{p}$, should be higher than that of the film layer, $n_\mathrm{c}$. Therefore high-index glasses, crystals like GGG (Gadolinium Gallium Garnet), SrTiO${}_3$ (strontium titanate), and TiO${}_2$ (rutile) are used often. In the case of rutile, the birefringence has to be taken into account.

Due to the phase-matching condition it is possible to excite the waveguide modes selectively by simply choosing a proper angle of incidence for the incoming laser beam. The laser beam is said to be in \emph{synchronous direction} then. 
The power is transferred from the prism into the waveguide along the coupling length. This can be seen by the observation of "black lines" at discrete angles in the output beam pattern\cite{Labadie:M-Lines}. Each mode has its own synchronous angle. The \emph{angular spectrum} of the intensity in the thin film exhibits a maximum at a finite number of these incident angles, related to the corresponding modes via \eqref{eq:Theo:mode_disp_eq}.

A major difficulty in implementing our approach is the typically narrow bandwidth due to the wavelength dependence of the mode dispersion equation \eqref{eq:Theo:mode_disp_eq}, as it depends both directly on the wavelength, and also indirectly through the wavelength dependence of the refractive indices. Therefore, quasi-non-dispersive materials like silicon or germanium have to be used in broadband implementations.

\section{The PHOTO Experiment}
The newly developed PHOTO (\textbf{P}rism-coupling, \textbf{H}igh-contrast, \textbf{O}ptical \textbf{T}unneling experiment using integrated \textbf{Optics}) Experiment is an adaption of the $m$-lines setup presented by several groups in the past\cite{feldman1983index, Labadie:M-Lines,Comparison, Ulrich:73, FTIR_paper, Tien:71, Tien:69} for high-contrast imaging purposes bearing in mind that
\begin{figure}[b]
	\begin{center}
		\begin{tabular}{c}
				\includegraphics[scale=.9]{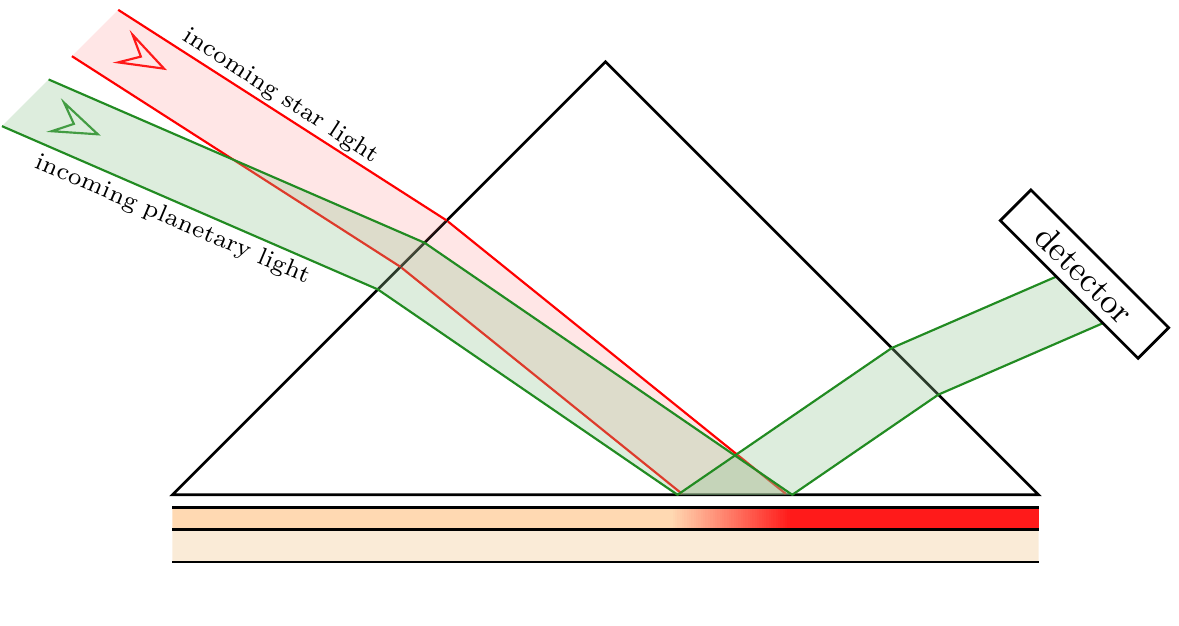}
		\end{tabular}
	\end{center}
	\caption[example]
	{ \label{fig:PHOTO_Principle} 
	Principle of the PHOTO Experiment. The star light is coupled into one mode of the waveguide while the planetary light is totally internally reflected at the base. It can be analyzed at the output facet of the prism. The angular separation of the star and the planet has been magnified by a telescope.}
\end{figure}
more than 200 extra-solar planets have been discovered while the direct detection still remains extremely challenging. The search for planets in the habitable zones of sun-like stars is following an ancient philosophical tradition and was always of great interest to the public. The advantageous contrast ratio between stars and planets using mid-infrared spectroscopy could be used to characterize the atmosphere of detected exoplanets\cite{Darwin-paper}.

\subsection{Starlight Rejection}
Various measurement methods like the radial velocity method, the transit method, or gravitational microlensing have been used for detecting exoplanets indirectly. In general, only very large (considerably larger than Jupiter) exoplanets which are widely separated from their stars can be detected directly nowadays. Earth-like distant extra-solar planets are not directly detectable because they are only faint sources in the vicinity of the much brighter stars they orbit. Their light gets lost in the glare of the star. Therefore, selective starlight extinction is mandatory for direct detection and especially for characterization efforts of earth-like exoplanets.

The preferred metric for high-angular resolution techniques is called the Inner Working Angle (IWA) which is commonly defined as the $50\%$ off-axis throughput point of a starlight rejecting system (usually coronagraphs)\cite{mawet2012review}. It is often expressed in terms of the resolution element $\lambda/d$ where $\lambda$ and $d$ are the observing wavelength and the diameter of the telescope (i.e.~its primary mirror), respectively. For a given distance of a planetary system, a smaller IWA allows imaging closer to the central star.
Probing the inner region of distant solar systems is crucial to understand the evolution of rocky planets. Here the habitable zone of a vast majority of nearby planetary systems is located.
$\SI{1}{AU}$ at a distance of $\SI{10}{pc}$ corresponds to an IWA of $0.1^{\prime\prime}$ or $2 \cdot \lambda/d$ (for a telescope operating in the visible, having a diameter of $\SI{2}{m}$). The IWA corresponds to the smallest distance where a faint object, apart the star it orbits, can be detected.
In summary, small IWA capabilities will provide significant scientific gains. In addition, it also provides a substantial benefit for future space-based operations as it enables the usage of smaller telescopes.

If the telescope is operating at the diffraction limit, then this sets the minimal angular distance between the green and red beam (refer to Fig.~\ref{fig:PHOTO_Principle}), which in the best case would be as small as $1.0 \cdot \lambda/d$. This means that, if the extinction can be made as high as $100\%$, we could have a concept with an IWA as small as the theoretical diffraction limit. Although quite appealing on paper, this idea needs experimental confirmation. We will give first laboratory results in section \ref{scn:Laboratory}.

\subsection{Coronagraphs}
Coronagraphic systems are commonly used in rejecting starlight. However, only a dozen of existing and prospected coronagraphs are working at very small angles\cite{mawet2012review}, which translates into an IWA {\footnotesize $\lesssim$}~$4 \cdot \lambda/d$.

Coronagraphs are intended to block the disc of the Sun so that it is possible to see the corona. Essentially, a coronagraph produces an artificial solar eclipse. This can be thought of as an arrangement where the sky is imaged onto an intermediate focal plane containing an opaque spot and is then reimaged onto a photodetector. 

\noindent Basically there are two types of coronagraphs used:
\begin{description}
	\item \textbf{Band-limited coronagraph}: The band-limited coronagraph uses a mask that is designed to block light and manage diffraction effects caused by removal of the light.
	
	\item \textbf{Phase-mask coronagraph}: The phase-mask coronagraph uses a specially designed phase shifting mask to shift the phase of the incoming star light in such a way that it interferes destructively with itself.
\end{description}

\noindent However, coronagraphs suffer from several disadvantages as several techniques in eliminating scattered light by diffraction have to be implemented, e.g.~the so-called Lyot stops are used to block the outer edge of the telescope pupil -- the angular resolution of the telescope is reduced to reduce the amount of flare caused by diffraction.

In contrast to coronagraphy, the PHOTO Experiment can deliver selective but meanwhile robust star-light extinction without affecting nearby (faint) sources at all as it is based on the phase-matching condition known from integrated optics: a very narrow angular interval of incident light is getting coupled into a waveguide. By using a well defined coupling gap below the prism, unity coupling efficiency, resulting in excellent starlight extinction, may be achieved. It provides a small setup that can be retrofitted to telescopes easily. The problem of having scattered light due to diffraction at the mask of the coronagraph is non-existing in the PHOTO Experiment.

\subsection{Integration into a Telescope}
\begin{figure}[h]
	\begin{center}
		\begin{tabular}{c}
			\includegraphics[scale=1.1]{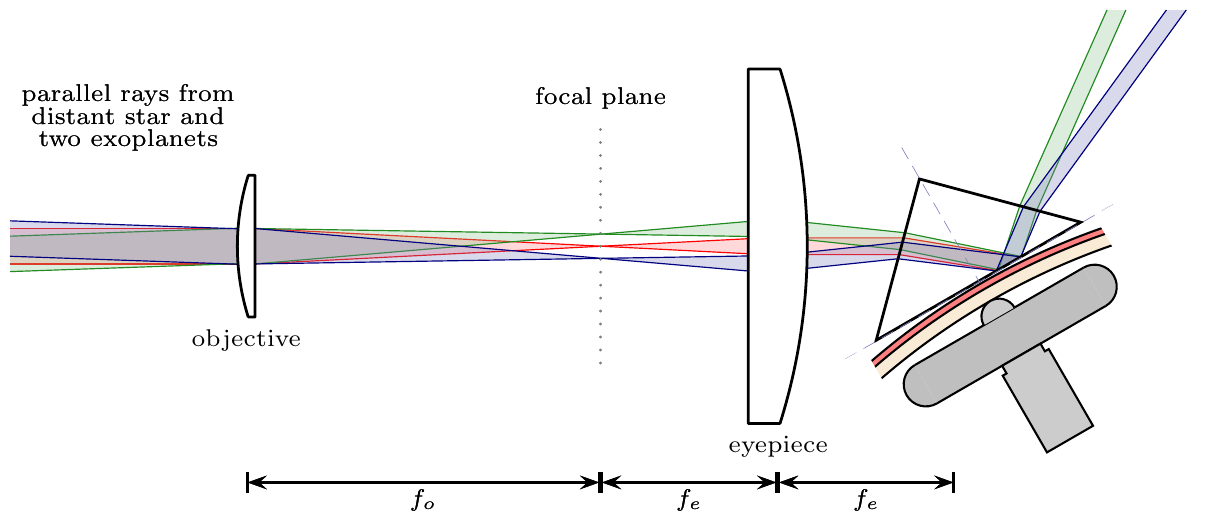}
		\end{tabular}
	\end{center}
	\caption[example]
	{ \label{fig:Telescope_incl_BELIEVE} 
	(Color online) PHOTO Experiment used to remove the bright light coming from the star (red ray). The sources can be selected by rotating the prism stage. The coupling gap underneath the prism has been bent to optimize coupling.}
\end{figure}

\begin{figure}[h]
\centering
	\subfigure[]{\includegraphics[width=0.24\textwidth]{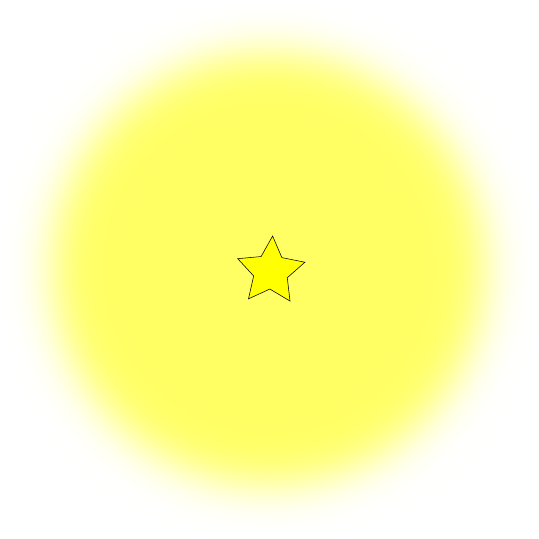}}
	\subfigure[]{\includegraphics[width=0.24\textwidth]{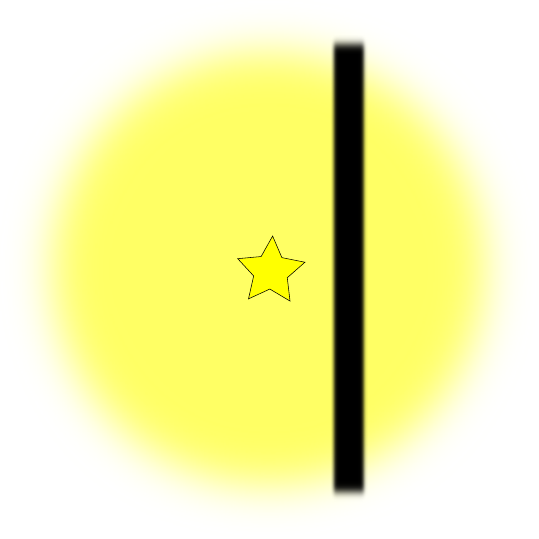}}
	\subfigure[]{\includegraphics[width=0.24\textwidth]{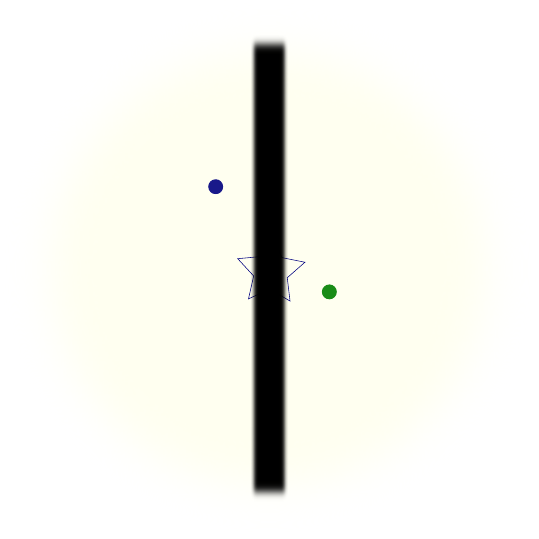}}
	\caption{\label{fig:PrincipleOfOperation} PHOTO Experiment (a) Direct image of the stellar system without the PHOTO experiment. (b) Tuning of the ``black line'' (angular direction that is coupled into the waveguide) towards the angular position of the bright central star. The shown alignment does not fulfil the phase-matching condition \eqref{eq:Theo:phase-matching_condition}. (c) The central star is properly aligned. The stellar light is coupled out into the waveguide. Consequently, the two faint companions are no longer covered by the stellar light and can be detected at the output facet of the prism as sketched in Figure \ref{fig:Telescope_incl_BELIEVE}.}
\end{figure}

The integration of the PHOTO Experiment is shown at a Keplerian telescope system that uses a convex lens as the eyepiece so that the emerging light is converging. Considerably high magnifications can be achieved with this setup. The PHOTO Experiment is located at the eye point where the incoming rays are intersecting (Fig.~\ref{fig:Telescope_incl_BELIEVE}). This implementation can be adapted for most other kinds of telescopes. The principle of operation is shown in Figure \ref{fig:PrincipleOfOperation}.

\subsection{Second Mode of Operation}
The PHOTO Experiment offers a second mode of operation, as, in contrast to coronagraphic methods, the light that is removed from the final image is not lost. In fact, this particular light is further processable, since it is coupled into the waveguide highly efficiently so that spectral analysis can be performed on this light. This could be done inside the waveguide itself by implementing systems like recently presented state-of-the-art microspectrometer\cite{wolffenbuttel2004state} leading to a highly integrated and compact instrument.

\section{Experimental Setup}
\begin{figure}[h]
	\begin{center}
		\begin{tabular}{c}
			\includegraphics[scale=1]{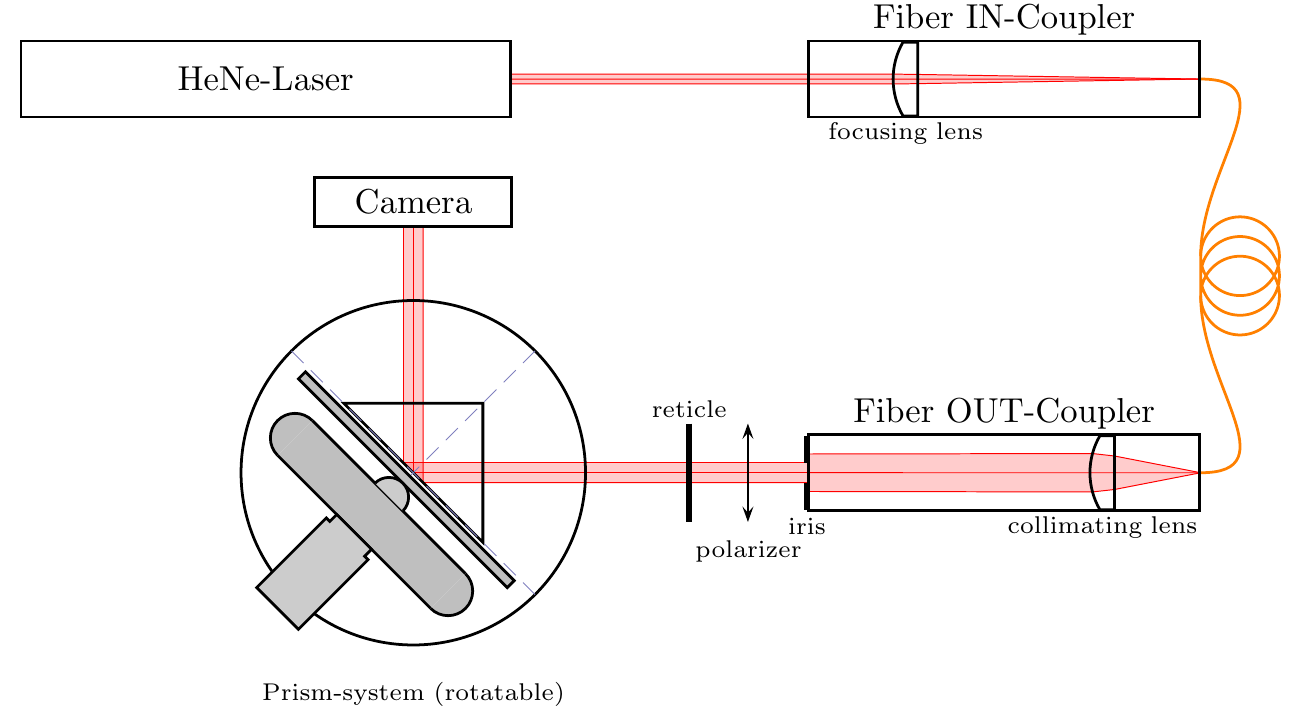}
		\end{tabular}
	\end{center}
	\caption[example]
	{ \label{fig:PHOTO_Setup_NEW} 
	The experimental setup used in our work.}
\end{figure}

Our experimental setup (shown in Figure \ref{fig:PHOTO_Setup_NEW}) involves a HeNe-laser whose beam is focused to meet the core of a single-mode fiber. At the other end of the fiber, the beam is collimated, and -- in contrast to the original laser beam -- spatially even and mode selected. The emerging beam has been adjusted to have a negligible divergence, leading to a very good approximation of plane waves like the ones coming from distant objects in the sky.
A fifteen-blade iris diaphragm and a linear film polarizer (antireflective (AR) coated) are used to control aperture and polarization of the emerging beam. The polarizer allows to select the desired polarization to probe either TE or TM modes.\enlargethispage{1\baselineskip}

The collimated beam reaches the base of the prism where it is totally internally reflected. We have placed a waveguide parallel to the base of the prism. To achieve an adjustable and continuous spacing between prism and waveguide, we use an engineered support similar to the one used by Ulrich and Torge\cite{Ulrich:73}. The waveguide is slightly bent and pushed towards the prism using a micrometer drive. We have modified the prism mount in such a way that it can be displaced linearly in $x$ and $y$ directions. The whole prism coupler assembly is mounted on a computer-controlled high-precision rotational stage.

The illuminated area of the prism base can be adjusted by displacing either the prism stage or the incident beam relative to each other. This can be used to continuously change the point of incidence on the prism base. In this configuration, high coupling efficiency can be achieved using a fairly simple setup.

\section{First Laboratory results} \label{scn:Laboratory}
\begin{figure}[h]
	\centering %
	\def\svgwidth{0.8\textwidth} %
	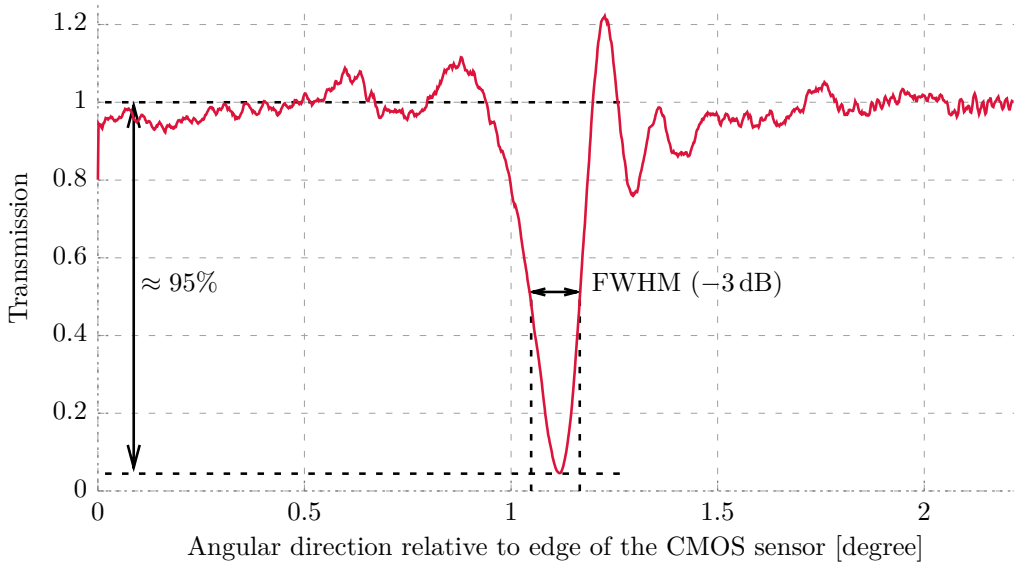
	\caption{ \label{fig:linewidth}
	The recorded intensity profile of the homogeneous incident laser beam measured at the output prism facet. The peaks above the value of 1 are due to decoupled light from the waveguide.
	}
\end{figure}

At the synchronous angle a reduction in the return beam intensity is measured. From the fractional change in the beam intensity one can directly get the coupling efficiency.
Our proof-of-concept laboratory setup showed that one can routinely achieve a coupling efficiency (extinction) of up to $\SI{95.0(2)}{\percent}$ even when using non-optimized waveguides (see Fig.~\ref{fig:linewidth}). The term \emph{non-optimized} refers in our case to the fact that neither the surfaces of the waveguides nor the prism base have been processed in any way. Earlier publications specific to optical tunneling suggest that nearly unity coupling efficiency is possible by slightly deforming the waveguide using mechanical stress\cite{Ulrich:71} or by designing the waveguide surface in a specific manner as has been described elsewhere\cite{thesisDerigs,Ulrich:73}.
Therefore, the PHOTO Experiment is expected to allow much higher coupling efficiencies when using designed waveguides.

The Full Width at Half Maximum (FWHM) linewidth where the intensity is reduced by 3 dB has been measured to be about $\SI{0.108(2)}{\degree} \simeq \SI{6.44(15)}{\arcmin}$. Although this appears quite unspectacular, one has to keep in mind that the PHOTO Experiment will be placed at the backside focal plane of the telescope. Therefore the linewidth has to be divided by the angular magnification of the telescope. When using telescopes with high angular magnifications, like the Very Large Telescope (VLT), the on-sky linewidth\cite{thesisDerigs} could be as small as
\begin{equation}
	\theta_\mathrm{sky} \approx \SI{0.01}{\arcsecond} \ .
\end{equation}
Therefore, operation at the diffraction-limit seems possible. Of course, the practical realization has to be studied in more detail in future works.

\section{Summary}
A theoretical description of optical tunneling and waveguide theory has been given where we showed that despite total internal reflection we still have an evanescent field that can be used to couple light into a waveguide. This is the underlying principle of our work.
First laboratory results show that diffraction limited operation seems to be possible using todays instruments. The measurement of exoplanetary spectral features may even be performed when unity coupling efficiency is somehow not achievable as the PHOTO Experiment offers a promising second mode of operation.
The extinction ratio, which is nothing else but the coupling efficiency, has been found to be very high (up to $\approx 95\%$) even when using non-optimized waveguides.

The state-of-the-art Hybrid Band-Limited coronagraph currently\cite{mawet2012review} holds the world record of the deepest and broadest (in terms of bandwidth) contrast level that can be demonstrated in the lab: down to roughly $3\cdot \lambda/d$. Using the PHOTO experiment it might be possible to achieve an IWA of almost $1 \cdot \lambda/d$ as our approach does not suffer from the diffraction problems encountered when using coronagraphs.

Regardless which mode of operation will be used, the measured atmospheric spectra could be examined for biomarker molecules that are indicative of biological processes. This offers perhaps the best opportunity to make first detection of extraterrestrial life.

\acknowledgments     
 
D.Derigs acknowledges support through the \emph{Bonn-Cologne Graduate School of Physics and Astronomy} (BCGS).

\bibliography{report}   
\bibliographystyle{spiebib}   

\end{document}

%% file: img/disp_eq.pdf_tex
\begingroup%
  \makeatletter%
  \providecommand\color[2][]{%
    \errmessage{(Inkscape) Color is used for the text in Inkscape, but the package 'color.sty' is not loaded}%
    \renewcommand\color[2][]{}%
  }%
  \providecommand\transparent[1]{%
    \errmessage{(Inkscape) Transparency is used (non-zero) for the text in Inkscape, but the package 'transparent.sty' is not loaded}%
    \renewcommand\transparent[1]{}%
  }%
  \providecommand\rotatebox[2]{#2}%
  \ifx\svgwidth\undefined%
    \setlength{\unitlength}{480bp}%
    \ifx\svgscale\undefined%
      \relax%
    \else%
      \setlength{\unitlength}{\unitlength * \real{\svgscale}}%
    \fi%
  \else%
    \setlength{\unitlength}{\svgwidth}%
  \fi%
  \global\let\svgwidth\undefined%
  \global\let\svgscale\undefined%
  \makeatother%
  \begin{picture}(1,0.7)%
    \put(0,0){\includegraphics[width=\unitlength]{disp_eq.pdf}}%
    \put(0.106,0.0885){\makebox(0,0)[rb]{\smash{-1}}}%
    \put(0.106,0.18566667){\makebox(0,0)[rb]{\smash{-0.5}}}%
    \put(0.106,0.283){\makebox(0,0)[rb]{\smash{0}}}%
    \put(0.106,0.38016667){\makebox(0,0)[rb]{\smash{0.5}}}%
    \put(0.106,0.4775){\makebox(0,0)[rb]{\smash{1}}}%
    \put(0.106,0.57466667){\makebox(0,0)[rb]{\smash{1.5}}}%
    \put(0.11983333,0.0585){\makebox(0,0)[b]{\smash{50}}}%
    \put(0.22816667,0.0585){\makebox(0,0)[b]{\smash{55}}}%
    \put(0.33633333,0.0585){\makebox(0,0)[b]{\smash{60}}}%
    \put(0.44466667,0.0585){\makebox(0,0)[b]{\smash{65}}}%
    \put(0.553,0.0585){\makebox(0,0)[b]{\smash{70}}}%
    \put(0.66116667,0.0585){\makebox(0,0)[b]{\smash{75}}}%
    \put(0.7695,0.0585){\makebox(0,0)[b]{\smash{80}}}%
    \put(0.87766667,0.0585){\makebox(0,0)[b]{\smash{85}}}%
    \put(0.986,0.0585){\makebox(0,0)[b]{\smash{90}}}%
    \put(0.02933333,0.339){\rotatebox{90}{\makebox(0,0)[b]{\smash{$m$}}}}%
    \put(0.55283333,0.0135){\makebox(0,0)[b]{\smash{$\theta_\mathrm{c}$}}}%
    \put(0.217,0.59216667){\makebox(0,0)[b]{\smash{$\arcsin(n_\mathrm{sub}/n_\mathrm{c})$}}}%
    \put(0.45483333,0.6625){\makebox(0,0)[rb]{\smash{\wavelengthA (TE)}}}%
    \put(0.45483333,0.6325){\makebox(0,0)[rb]{\smash{\wavelengthA (TM)}}}%
    \put(0.788,0.6625){\makebox(0,0)[rb]{\smash{\wavelengthC (TE)}}}%
    \put(0.788,0.6325){\makebox(0,0)[rb]{\smash{\wavelengthB (TE)}}}%
  \end{picture}%
\endgroup%

%% file: img/linewidth_mod_norm.pdf_tex
\begingroup%
  \makeatletter%
  \providecommand\color[2][]{%
    \errmessage{(Inkscape) Color is used for the text in Inkscape, but the package 'color.sty' is not loaded}%
    \renewcommand\color[2][]{}%
  }%
  \providecommand\transparent[1]{%
    \errmessage{(Inkscape) Transparency is used (non-zero) for the text in Inkscape, but the package 'transparent.sty' is not loaded}%
    \renewcommand\transparent[1]{}%
  }%
  \providecommand\rotatebox[2]{#2}%
  \ifx\svgwidth\undefined%
    \setlength{\unitlength}{640bp}%
    \ifx\svgscale\undefined%
      \relax%
    \else%
      \setlength{\unitlength}{\unitlength * \real{\svgscale}}%
    \fi%
  \else%
    \setlength{\unitlength}{\svgwidth}%
  \fi%
  \global\let\svgwidth\undefined%
  \global\let\svgscale\undefined%
  \makeatother%
  \begin{picture}(1,0.5625)%
    \put(0,0){\includegraphics[width=\unitlength]{linewidth_mod_norm.pdf}}%
    \put(0.0795,0.066375){\makebox(0,0)[rb]{\smash{0}}}%
    \put(0.0795,0.1415){\makebox(0,0)[rb]{\smash{0.2}}}%
    \put(0.0795,0.216625){\makebox(0,0)[rb]{\smash{0.4}}}%
    \put(0.0795,0.29175){\makebox(0,0)[rb]{\smash{0.6}}}%
    \put(0.0795,0.366875){\makebox(0,0)[rb]{\smash{0.8}}}%
    \put(0.0795,0.442125){\makebox(0,0)[rb]{\smash{1}}}%
    \put(0.0795,0.51725){\makebox(0,0)[rb]{\smash{1.2}}}%
    \put(0.089875,0.043875){\makebox(0,0)[b]{\smash{0}}}%
    \put(0.288,0.043875){\makebox(0,0)[b]{\smash{0.5}}}%
    \put(0.48625,0.043875){\makebox(0,0)[b]{\smash{1}}}%
    \put(0.684375,0.043875){\makebox(0,0)[b]{\smash{1.5}}}%
    \put(0.882625,0.043875){\makebox(0,0)[b]{\smash{2}}}%
    \put(0.022,0.30675){\rotatebox{90}{\makebox(0,0)[b]{\smash{Transmission}}}}%
    \put(0.52925,0.010125){\makebox(0,0)[b]{\smash{Angular direction relative to edge of the CMOS sensor [degree]}}}%
    \put(0.131125,0.264875){\makebox(0,0)[lb]{\smash{$\approx 95\%$}}}%
    \put(0.56375,0.264875){\makebox(0,0)[lb]{\smash{FWHM ($\SI{-3}{dB}$)}}}%
  \end{picture}%
\endgroup%